\documentclass[aps,12pt,preprintnumbers,nofootinbib,superscriptaddress]{revtex4}
\usepackage{graphicx}
\usepackage{amssymb,amsmath}
\def\Dslash{D\hskip-0.65em /}
\def\Aslash{A\hskip-0.65em /}

\begin{document}

\title{\Large{\bf Aspects of Electrodynamics Modified by Some Dimension-five Lorentz Violating Interactions } }

\author{Shan-quan Lan}
\author{Feng Wu}
\email[Electronic address: ]{fengwu@ncu.edu.cn}
\affiliation{%
Department of Physics, Nanchang University,
330031 Nanchang, China}

\date{\today}

\begin{abstract}
Assuming Lorentz symmetry is broken by some fixed vector background, we study the spinor electrodynamics modified by two dimension-five Lorentz-violating interactions between fermions and photons. The effective polarization and magnetization are identified from the modified Maxwell equations, and the theoretical consequences are investigated. We also compute the corrections to the relativistic energy levels of hydrogen atom induced by these Lorentz-violating operators in the absence and presence of uniform external fields in first-order perturbation theory. We find that the hydrogen spectrum is insensitive to the breakdown of Lorentz boost symmetry.  
\end{abstract}
\pacs{} 
\maketitle 
\newpage

\date{\today}

\section{Introduction}
The breaking of Lorentz symmetry is an extensively studied topic. Although no departure from Lorentz invariance has yet been detected experimentally, there is no reason to believe that Lorentz invariance would be intact at all energies. As a matter of fact, there are reasons to suspect its exactness in the context of string theory. For example, the potential instability of string vacuum would induce spontaneous breakdown of Lorentz symmetry \cite{Samuel}. Also, high-energy field theories constructed on a Moyal space, viewed as low-energy effective theories from open string theory with a constant NS-NS $B$ field, explicitly spoil Lorentz invariance \cite{S-W}.

One direction in the study of Lorentz violation is to regard Lorentz symmetry breaking as a possible extension of the standard model in particle physics. The first work in this context is the investigation of the Carroll-Field-Jackiw term \cite{CFJ}, which has inspired numbers of studies in recent years. Without the criterion of Lorentz symmetry, one may construct new additive terms to the minimal standard model. Note that since $CPT$ invariance is necessary but not sufficient for Lorentz invariance of an interacting quantum field theory \cite{CPT}, Lorentz violating (LV) terms can be either $CPT$ even or $CPT$ odd. LV terms of renormalizable dimensions have been systematically constructed in \cite{Colladay}, known as the standard model extension, and many related issues have been discussed \cite{fermion, radiative, CPT1, ODD, EVEN}. However, experimental data put very stringent constraints on the renormalizable LV terms and indicate that they must be extremely small. To avoid the subtle fine-tuning problem \cite{FT}, we assume that the symmetry of the underlying theory prohibits the generation of the renormalizable LV operators and explore the nonrenormalizable LV operators in this paper.

Renormalizability was considered to be an axiom when constructing the standard model. Quantum corrections to a renormalizable theory will generate UV divergences only to operators whose mass dimensions are less than five and this fact assures the predictiveness of the theory. However, a modern point of view is that reliable predictions could still be made from a nonrenormalizable theory within the framework of effective field theories. Thus, higher-dimensional LV operators should also be considered, and several studies involving dimension-five LV operators have been carried out \cite{HD1, constraint, HD2, Hydrogen, HD3}.

In this paper, assuming a fixed vector background $v^{\mu}$ to be the only source that induces the breaking of Lorentz symmetry, we shall consider the spinor electrodynamics modified by some dimension-five LV interactions between fermions and photons. Higher-dimensional interactions are best classified in terms of derivative expansion. Dimension-five interactions are quadratic in gauge covariant derivative $D^{\mu}$, which is given by $D^{\mu}=\partial^{\mu} + i A^{\mu}$. Recall that $[D^{\mu} , D^{\nu} ]=i F^{\mu\nu}$, where $F^{\mu\nu}$ is the electromagnetic tensor. If we restrict ourself to consider only terms linear in the vector background $v^{\mu}$ and the photon field $A^{\mu}$, then the most general dimension-five interactions are of the form 
\begin{equation}
\mathcal{L}_{v}= v^{\mu} \overline{\Psi}K_{\mu\nu\alpha\beta}  \gamma^{\nu} \Psi F^{\alpha \beta} \label{LLV}
\end{equation} 
where $K_{\mu\nu\alpha\beta}={1\over 2}\left( a_{1} + b_{1} \gamma^{5} \right) \epsilon_{\mu\nu\alpha\beta} + \left( a_{2} + b_{2} \gamma^{5} \right) g_{\mu\alpha}g_{\nu\beta}$ with $a_{i}$ and $b_{i}$ being dimensionless constants. The factor $1/2$ in $K_{\mu\nu\alpha\beta}$ is introduced for later convenience. One can easily see that the mass dimension of the background vector $v^{\mu}$ is  $[v^{\mu}]=-1$. Note that each term in~(\ref{LLV}) violates $CPT$. While $a_{i}$ terms preserve $C$ parity (and thus violate $PT$), $b_{i}$ terms violated it (and thus preserve $PT$). Constraints from the electric dipole moments of paramagnetic atoms put very stringent limits on $b_{i}$ terms \cite{constraint}. Therefore, we will not discuss $C$-violating terms in this paper and simply set $b_{i}=0$ from now on.

The modified QED, after rescaling the background $v^{\mu}$ by absorbing the parameter $a_{1}$, then reads
\begin{equation}
\mathcal{L}=-{1\over 4 e^2} F_{\mu\nu} F^{\mu\nu} + \overline{\Psi} \left( i \Dslash -m -\gamma^{\mu}v^{\nu} \left( \tilde{F}_{\mu\nu} + a F_{\mu\nu} \right) \right) \Psi  \label{L}
\end{equation}    
where $\tilde{F}_{\mu\nu} $ is the dual electromagnetic tensor, $\tilde{F}_{\mu\nu}\equiv {1\over 2}\epsilon_{\mu\nu\alpha\beta}F^{\alpha\beta}$. The Lorentz symmetry $SO(3,1)$ is broken by the irrelevant dimension-five operators to its subgroup $SO(2)$, which admits the background vector $v^{\mu}$ as an invariant tensor. At low energies, effects due to nonrenormalizable couplings are suppressed at least by powers of $1/M $, with $M$ being some fundamental large mass scale in the underlying theory. In the limit $M\rightarrow \infty$, the symmetry of the Lagrange density~(\ref{L}) is enhanced to the Lorentz group, along with spacetime translations.   

The two LV terms in~(\ref{L}) have been considered in Refs.~\cite{HD2, Hydrogen, HD3}. The crucial difference between the existing works related to these two terms and the Lagrange density~(\ref{L}) considered in this paper is that the dimensionless coupling constant $e$ in~(\ref{L}) is the unique gauge coupling constant determining the strength of the electromagnetic interaction. Thus, different from other works, we restrict our consideration to the case where electrically neutral particles will not interact with photons at tree level. This can be seen even more transparently by letting $A^{\mu} \rightarrow e A^{\mu}$ so that~(\ref{L}) becomes 
\begin{equation}
\mathcal{L}=-{1\over 4} F_{\mu\nu} F^{\mu\nu} + \overline{\Psi} \left( i \Dslash -m \right) \Psi -j^{\mu}  v^{\nu} \left(\tilde{F}_{\mu\nu} + a F_{\mu\nu}\right)   \label{L2}
\end{equation}      
where the gauge covariant derivative now takes the form $D^{\mu}=\partial^{\mu} + ie A^{\mu}$ and the $4$-vector $j^{\mu}\equiv e \overline{\Psi} \gamma^{\mu} \Psi = ( \rho, \vec{j})$ is the current density. Apparently $\mathcal{L}$ reduces to the free theory for neutral particles.

The rest of the paper is organized into three parts. In Sec. II, we examine the QED modified by the dimension-five LV operator, $j^{\mu} v^{\nu} \tilde{F}_{\mu\nu}$. The theoretical consequences of the modified Maxwell and Dirac equations are studied. In particular, we compute the corrections to the hydrogen spectrum by applying the perturbation theory to the exactly solved Dirac equation. To our knowledge, the corrections to the hydrogen spectrum induced by the LV operator $ j^{\mu} v^{\nu} \tilde{F}_{\mu\nu}$ were calculated only in the nonrelativistic limit in the literature \cite{Hydrogen}. The effect on the spectral lines of hydrogen atom due to the presence of a static external electric field and a static external magnetic field is also considered. In Sec. III, we present similar analysis on the QED modified by another dimension-five LV operator, $j^{\mu} v^{\nu} F_{\mu\nu}$. We give our conclusions in the final section.  

\section{Model I}
Our starting point is the following modified QED Lagrange density:
\begin{equation}
\mathcal{L}_{1}=-{1\over 4} F_{\mu\nu} F^{\mu\nu} + \overline{\Psi} \left( i \Dslash -m -e\gamma^{\mu}v^{\nu}  \tilde{F}_{\mu\nu}   \right) \Psi .  \label{L1}
\end{equation} 
The field equations derived from $\mathcal{L}_{1}$ read\footnote{The convention for the metric in this paper has the signature$(+,-,-,-)$}
\begin{equation}
\partial_{\nu}F^{\nu\mu} = j^{\mu} + \epsilon^{\mu\nu\alpha\beta}v_{\beta}\partial_{\nu} j_{\alpha}. \label{L1eom}
\end{equation} 
The continuity equation $\partial_{\mu} j^{\mu} =0$ follows from~(\ref{L1eom}) as a result of gauge symmetry. The field equations~(\ref{L1eom}) can be rewritten in terms of components as the familiar form of inhomogeneous Maxwell equations. Together with the homogeneous Maxwell equations coming from the gauge invariance of the system, we have
\begin{eqnarray}
\overrightarrow{\nabla}\cdot \overrightarrow{B} =0, \\
{\partial \overrightarrow{B} \over \partial t} + \overrightarrow{\nabla}\times \overrightarrow{E}=0,\\
\overrightarrow{\nabla}\cdot \overrightarrow{D} =\rho, \\
\overrightarrow{\nabla}\times \overrightarrow{H}-{\partial \overrightarrow{D} \over \partial t} =\overrightarrow{j}.
\end{eqnarray}
Here the effective displacement field $\overrightarrow{D}$ and the effective magnetic field $\overrightarrow{H}$ are defined as $\overrightarrow{D}=\overrightarrow{E}+\overrightarrow{P}$ and $\overrightarrow{H}=\overrightarrow{B}-\overrightarrow{M}$, respectively, where the effective polarization $\overrightarrow{P} $ and the effective magnetization $\overrightarrow{M}$, defined by $\overrightarrow{P}=(\overrightarrow{j}\times\overrightarrow{v}) $ and   $\overrightarrow{M}=(\rho\overrightarrow{v}-v^{0}\overrightarrow{j} )$, respectively, are the components of the rank-$2$ object $M_{\mu\nu} \equiv j_{[\mu} v_{\nu]}$:
\begin{equation}
M_{\mu\nu}=\left( \begin{array}{cccc}
0     & -M^{1} & -M^{2} & -M^{3} \\
M^{1} & 0      & P^{3}  & -P^{2} \\
M^{2} & -P^{3} &  0     &  P^{1}\\
M^{3} &  P^{2} & -P^{1} &  0  \end{array} \right).
\end{equation} 
Lorentz symmetry of the Maxwell equations is spoiled by the nonzero effective polarization or the nonzero effective magnetization. In the presence of stationary sources (such that $\vec{\nabla}\cdot \vec{j}=0$), the solution of the gauge field $A^{\mu}=(\phi (\vec{r}), \vec{A}(\vec{r}) )$ is 
\begin{equation}
\phi(\vec{r})= {1\over 4 \pi} \int d^{3} r' { \rho(\vec{r}\,') - \vec{\nabla} \cdot \vec{P}(\vec{r}\,') \over \vert \vec{r} -\vec{r}\,' \vert } 
\end{equation} 
and
\begin{equation}
\vec{A}(\vec{r})= {1\over 4 \pi} \int d^{3} r' { \vec{j}(\vec{r}\,') + \vec{\nabla}\,' \times \vec{M}(\vec{r}\,') \over \vert \vec{r} -\vec{r}\,' \vert } .
\end{equation} 
Consequently, one can see that a static electric field can arise from stationary and neutral sources ($\rho=0$) as long as the effective polarization is not divergence-free. Also, even for the steady and irrotational current density (such that both the divergence and the curl of $\vec{j}$ vanish), a nonvanishing magnetic field $\vec{B}$ may still arise from $\vec{B}= \vec{\nabla}\times \vec{A}$ with $\vec{A}$ given by 
\begin{equation}
\vec{A}(\vec{r})= {1\over 4 \pi} \int d^{3} r' { ( \vec{\nabla}\,' \rho(\vec{r}\,') )\times \vec{v} \over \vert \vec{r} -\vec{r}\,' \vert } .
\end{equation} 

We now switch to the fermion sector in~~(\ref{L1}). The equation of motion for the fermion $\Psi$ following from $\mathcal{L}_{1}$ is 
\begin{equation}
\left(i \Dslash -m - e \gamma^{\mu}v^{\nu} \tilde{F}_{\mu\nu} \right) \Psi=0 .
\end{equation} 
Multiplying on the left by the Dirac matrix $\gamma^{0}$, we can identify the Hamiltonian operator of one-particle quantum mechanics:
\begin{equation}
H=\gamma^{0} \left(\vec{\gamma}\cdot \vec{p} + e\Aslash + m + ev^{0} \vec{\gamma}\cdot \vec{B} -e \gamma^{0} \vec{v}\cdot \vec{B} -e \vec{\gamma}\cdot (\vec{v}\times\vec{E})  \right)\\
=H_{0} +\delta H \label{H1}
\end{equation}
where $H_{0}= \gamma^{0} (\vec{\gamma}\cdot \vec{p} + e\Aslash + m )$ is the Dirac Hamiltonian and $\delta H= e\gamma^{0}( v^{0} \vec{\gamma}\cdot \vec{B} - \gamma^{0} \vec{v}\cdot \vec{B} - \vec{\gamma}\cdot (\vec{v}\times\vec{E})  )$ is the LV perturbation. It is well known that hydrogen atom can be solved exactly in Dirac's theory and the fine structure of the hydrogen spectrum comes out naturally from it. Using degenerate perturbation theory, we are able to compute the first-order correction to the hydrogen spectrum induced by $\delta H$.

Since the degenerate unperturbed states are the stationary state vectors $\vert n,j,l,m_{j}\rangle $ of the Dirac Hamiltonian $H_{0}$ for a fixed $n$ and $j$, in the absence of external fields we need to calculate the following matrix elements of the perturbation:  

\begin{equation}
\langle n,j,l\,', m_{j}'  \vert \delta H \vert n,j, l, m_{j} \rangle=-e \langle n,j,l\,', m_{j}'  \vert \gamma^{0} \vec{\gamma}\cdot (\vec{v} \times \vec{E}) \vert n,j, l, m_{j} \rangle \label{MA}
\end{equation}
where the Coulomb field $\vec{E}$ is given by $\vec{E} = - {e\over 4 \pi} {\hat{r} \over r^2}$. The term $-e \gamma^{0}\vec{\gamma}(\vec{v}\times \vec{E})$ in $\delta H$ originates from the $CP$-even operator $j^{i}v^{k}\tilde{F}_{ik}$ in $\mathcal{L}_{1}$. We note in passing that the energy shifts are independent of the time component $v^{0}$ of the background vector, indicating that the hydrogen spectrum is insensitive to the breakdown of invariance under Lorentz boosts. It is easy to show that the matrix elements~(\ref{MA}) of the perturbation between state vectors with different eigenvalues for the square of the orbital angular momentum $L^2$ or the $z$ component $J_{z}$ of the total angular momentum all vanish. Indeed, by judiciously choosing a coordinate system such that $\vec{v}=\vert \vec{v} \vert \hat{z}$, we have, in Dirac representation,
 \begin{eqnarray}
& &[J_{z}, \gamma^{0}\vec{\gamma} \cdot (\vec{v} \times \vec{E})]\propto [-i{\partial \over \partial \phi} \mathrm{1} +{1\over 2} \left( \begin{array}{cc}
\hat{\sigma}_{z}    & 0 \\
0 & \hat{\sigma}_{z}     \\
\end{array} \right), \left( \begin{array}{cc}
0    & \mathrm{sin}\phi \hat{\sigma}_{x}-\mathrm{cos} \phi \hat{\sigma}_{y}   \\
 \mathrm{sin}\phi \hat{\sigma}_{x}-\mathrm{cos} \phi \hat{\sigma}_{y}&0    \\
\end{array}\right)] \nonumber \\
& &=-i\left( \begin{array}{cc}
0    & \mathrm{cos}\phi \hat{\sigma}_{x}+\mathrm{sin} \phi \hat{\sigma}_{y}   \\
 \mathrm{cos}\phi \hat{\sigma}_{x}+\mathrm{sin} \phi \hat{\sigma}_{y}&0    \\
\end{array}\right)\nonumber \\
& &+{1\over2}\left( \begin{array}{cc}
0    & \mathrm{sin}\phi [\hat{\sigma}_{z},\hat{\sigma}_{x}]-\mathrm{cos} \phi [\hat{\sigma}_{z},\hat{\sigma}_{y}]   \\
\mathrm{sin}\phi [\hat{\sigma}_{z},\hat{\sigma}_{x}]-\mathrm{cos} \phi [\hat{\sigma}_{z},\hat{\sigma}_{y}] &0    \\
\end{array}\right)=0.
 \end{eqnarray}
Also, the unperturbed states $\vert n,j,l,m_{j}\rangle$ are simultaneous eigenstates of $H_{0}$ and $J_{z}$. It follows then that the matrix elements~(\ref{MA}) vanishes unless $l=l\,'$ and $m_{j}=m_{j}'$.
 
To evaluate the expectation value in the unperturbed state of the perturbation, we recall that the unperturbed wave functions in Dirac representation take the form
\begin{equation}
\langle x^{\mu} \vert n,j,l=j\pm{1\over 2},m_{j}\rangle = e^{-i{\epsilon}t}\binom{iF_{-}(\pm\kappa\vert r)\mathcal{Y}_{j,m_{j}}(j\pm{1\over 2},\frac{1}{2}\vert \hat{r})}{F_{+}(\pm\kappa\vert r)\mathcal{Y}_{j,m_{j}}(j\mp{1\over 2},\frac{1}{2}\vert \hat{r})}.
\end{equation}
Here the radial wave functions $ F_{\pm}(\kappa\vert r) $ are given by
\begin{equation}
 F_{\pm}(\kappa\vert r)={\mp}N_{\pm}(\kappa)(2{\mu}r)^{\gamma-1}e^{-{\mu}r}\{[\frac{({n'}+\gamma)m_{e}}{\epsilon}-\kappa]F(-{n'},2\gamma+1;2{\mu}r) {\pm}{n'}F(1-{n'},2\gamma+1;2{\mu}r)\} \end{equation}
where
\begin{eqnarray}
N_{\pm}(\kappa)&=&\frac{(2\mu)^{\frac{3}{2}}}{\Gamma(2\gamma+1)}\sqrt{\frac{(m_{e}\mp\epsilon)\Gamma(2\gamma+{n'}+1)}{4m_{e}\frac{({n'}+\gamma)m_{e}}{\epsilon}(\frac{({n'}+\gamma)m_{e}}{\epsilon}-\kappa){n'}!}},\nonumber\\
 \mu&=&\sqrt{(m_{e}-\epsilon)(m_{e}+\epsilon)},\nonumber\\
 \epsilon&=&\frac{m_{e}}{\sqrt{1+\frac{\alpha^{2}}{({n'}+\gamma)^{2}}}},\nonumber\\
  \gamma&=&\sqrt{(j+\frac{1}{2})^2-\alpha^{2}},\nonumber\\
 {n'}&=&n-\kappa,\nonumber\\
 \kappa&=&j+{1\over 2},
 \end{eqnarray} 
$\alpha $ is the fine structure constant given by $\alpha=e^2/4\pi$, and $m_{e}$ is the electron mass. The spin-angular functions $\mathcal{Y}_{j,m_{j}}(l,\frac{1}{2}\vert \hat{r}) $ are of the form
\begin{equation}
\mathcal{Y}_{j,m_{j}}(l,\frac{1}{2}\vert \hat{r})=\binom{(-1)^{l-j+\frac{1}{2}}\sqrt{\frac{l+\frac{1}{2}+(-1)^{l-j+\frac{1}{2}}m_{j}}{2l+1}}Y_{l}^{m_{j}-\frac{1}{2}}(\theta,\varphi)}{\sqrt{\frac{l+\frac{1}{2}+(-1)^{l-j+\frac{3}{2}}m_{j}}{2l+1}}Y_{l}^{m_{j}+\frac{1}{2}}(\theta,\varphi)}.
\end{equation}
It follows that
 \begin{eqnarray}
 &\,\,&-e\langle n,j,l\,', m_{j}'  \vert \gamma^{0} \vec{\gamma}\cdot (\vec{v} \times \vec{E}) \vert n,j, l, m_{j} \rangle =-\delta_{l l\,'} \delta_{m_{j} m_{j}'}e\langle n,j,l, m_{j} \vert \gamma^{0} \vec{\gamma}\cdot (\vec{v} \times \vec{E}) \vert n,j, l, m_{j} \rangle \nonumber \\
 &=&- \delta_{l l\,'} \delta_{m_{j} m_{j}'} \alpha \vert \vec{v} \vert{\int}d^{3}r {1\over r^2} \left(-iF_{-}(\pm \kappa\vert r)\mathcal{Y}_{j,m_{j}}^{\dagger}(j\pm\frac{1}{2},\frac{1}{2}\vert \hat{r}),F_{+}(\pm\kappa\vert r)\mathcal{Y}_{j,m_{j}}^{\dagger}(j\mp\frac{1}{2},\frac{1}{2}\vert \hat{r})\right)\left(\begin{array}{cc}0
& \overrightarrow{\sigma} \\ \overrightarrow{\sigma} & 0
\end{array}\right)\cdot\nonumber\\
&\,\,&( \sin{\theta}\sin{\varphi}\hat{x}
- \sin{\theta}\cos{\varphi}\hat{y})\binom{iF_{-}(\pm\kappa\vert r)\mathcal{Y}_{j,m_{j}}(j\pm\frac{1}{2},\frac{1}{2}\vert \hat{r})}{F_{+}(\pm\kappa\vert r)\mathcal{Y}_{j,m_{j}}(j\mp\frac{1}{2},\frac{1}{2}\vert \hat{r})} \nonumber\\
&=& \mp \delta_{l l\,'} \delta_{m_{j} m_{j}'}  \alpha \vert \vec{v}\vert \int
drd\cos{\theta}F_{-}(\pm\kappa\vert r)F_{+}(\pm\kappa\vert r)\sin{\theta}[\frac{(j-m_{j}+1)!}{(j+m_{j})!}P_{j+\frac{1}{2}}^{m_{j}-\frac{1}{2}}(\cos\theta)P_{j-\frac{1}{2}}^{m_{j}+\frac{1}{2}}(\cos\theta)\nonumber\\
&\,\,&+\frac{(j-m_{j})!}{(j+m_{j}-1)!}P_{j+\frac{1}{2}}^{m_{j}+\frac{1}{2}}(\cos\theta)P_{j-\frac{1}{2}}^{m_{j}-\frac{1}{2}}(\cos\theta)] \nonumber\\
&=&\mp \delta_{l l\,'} \delta_{m_{j} m_{j}'}   \alpha \vert \vec{v} \vert \int dr d\cos{\theta}F_{-}(\pm\kappa\vert r)F_{+}(\pm\kappa\vert r)[-\frac{(j-m_{j}+1)!}{(j+m_{j})!}\frac{(j-m_{j}+1)(j-m_{j})}{2j}(P_{j+\frac{1}{2}}^{m_{j}-\frac{1}{2}}(\cos\theta))^{2}\nonumber\\
&\,\,&+\frac{(j-m_{j})!}{(j+m_{j}-1)!}\frac{1}{2j}(P_{j+\frac{1}{2}}^{m_{j}+\frac{1}{2}}(\cos\theta))^{2}] \nonumber\\
&=&\mp \delta_{l l\,'} \delta_{m_{j} m_{j}'}  \alpha\vert\vec{v}\vert \frac{m_{j}(2j+1)}{j(j+1)}\int_{0}^{\infty}drF_{-}(\pm\kappa\vert r)F_{+}(\pm\kappa\vert r) \nonumber\\
&=&\pm\delta_{l l\,'} \delta_{m_{j} m_{j}'}\alpha\vert\vec{v}\vert \frac{m_{j}(2j+1)}{j(j+1)} N_{-}(\pm\kappa)N_{+}(\pm\kappa)  \int_{0}^{\infty}dr(2\mu{r})^{2\gamma-2}e^{-2\mu{r}}\nonumber\\
&\,\,&\{[\frac{({n'}+\gamma)m_{e}}{\epsilon}\mp\kappa]^{2}F^{2}(-{n'},2\gamma+1;2\mu{r})-{n'}^{2}F^{2}(1-{n'},2\gamma+1;2\mu{r})\}\nonumber\\
&=&\pm \delta_{l l\,'} \delta_{m_{j} m_{j}'} \frac{ \alpha \vert \vec{v}\vert(m_{e}^{2}-{\epsilon}^{2})^{\frac{3}{2}}m_{j}(2j+1)}{4 m_{e}^{2}j(j+1)\gamma(\gamma^{2}-{1\over4})(n+\gamma-j-{1\over2})}\nonumber \\
&&\left( (n+\gamma-j)\left((n+\gamma-j-{1\over2})m_{e} \mp (j+{1\over 2})\epsilon \right)- {\left( \left(n+\gamma-j-{1\over 2}\right)^2-\gamma^2    \right) \left(n+\gamma-j-1\right)\epsilon^2 \over   \left(n+\gamma-j-{1\over2}\right)m_{e} \mp (j+{1\over 2})\epsilon  }  \right )\label{dE}
 \end{eqnarray}
for $l=j\pm{1\over 2}$. In deriving the above result, we have used the following formula for confluent hypergeometric functions:
\begin{equation}
\int_{0}^{\infty}d\xi\, \xi^{2l-1} e^{-\xi} F^{2}(-n+l+1,2l+2;\xi) = {n\Gamma^{2}(2l+2) \Gamma(n-l) \over 4 l(l+{1\over 2})(l+1)\Gamma(n+l+1)}.
\end{equation}

Expanding~(\ref{dE}) in powers of the fine structure constant, we obtain the energy shifts produced by $\delta H$:
\begin{equation}
\delta E_{njlm_{j}}=- \vert \vec{v} \vert m_{e}^{2} \alpha^{4} { m_{j}  \over n^{3} j(j+1) (l+{1\over 2})}+ \mathrm{O} (\alpha^{6}).
\end{equation}
The degeneracy of the fine structure in $l$ and $m_{j}$ has been removed by the LV perturbation $\delta H$. Figure 1 shows the low-lying energy levels of the hydrogen atom. Note that the energy shifts $\delta E_{njlm_{j}}$ is of order $(m_{e} \vert \vec{v} \vert)m_{e} \alpha^{4}$, where $(m_{e} \vert \vec{v} \vert)$ is a dimensionless product. This is a tiny effect in comparison with the Lamb shift, which is of order $m_{e} \alpha^{5}$, and the hyperfine splitting, which is of order $(m_{e}/m_{p})m_{e}\alpha^{4}$ with $m_{p}$ being the mass of the proton, since the irrelevant LV operator $j^{\mu}v^{\nu} \tilde{F}_{\mu \nu}$ is highly suppressed by some large fundamental mass scale $M$ mentioned in the introduction. 
\begin{figure}[t]
\begin{center}
\includegraphics[width=15cm,clip=true,keepaspectratio=true]{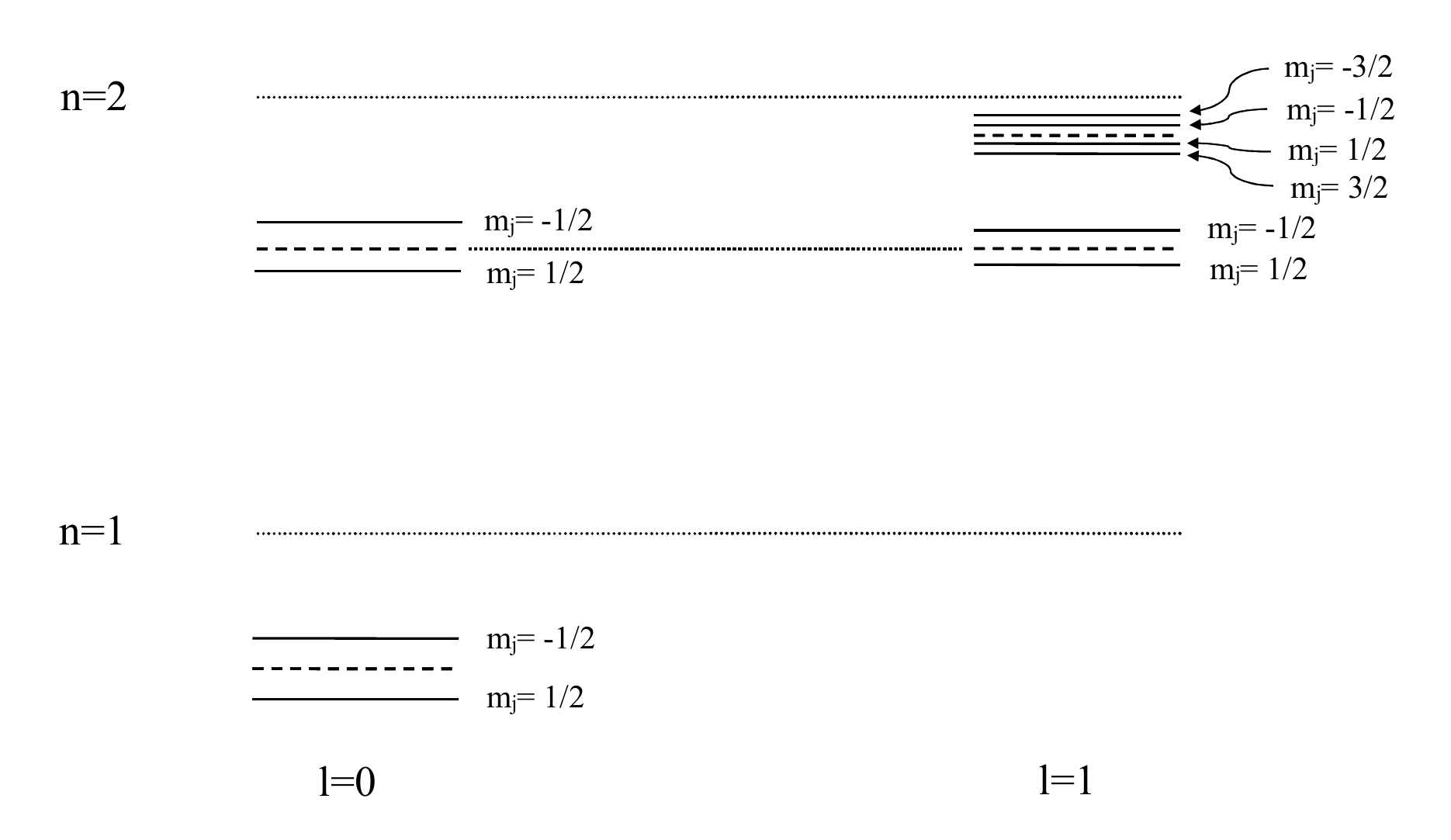}
\caption{\small The low-lying energy levels of the hydrogen atom, including the first-order LV correction (not to scale).}
\end{center}\label{loop}
\end{figure}
 
We are now in a position to consider the shift of the hydrogen energy levels in the presence of a uniform external magnetic field $\vec{B}_{ext}$, assuming that its strength is weak in comparison with the field produced by the proton. The unperturbed Hamiltonian is taken to be the Dirac Hamiltonian $H_{0}$ in the absence of the magnetic field. The term $e\gamma^{0}\vec{\gamma}\cdot \vec{A}$ in $H_{0}$ is thus treated as a perturbation and responsible for the well-known Zeeman effect in the nonrelativistic limit. With the LV perturbation $\delta H$, we now also need to consider the effect induced by the terms $e(v^{0} \gamma^{0}\vec{\gamma}\cdot \vec{B}_{ext}- \vec{v}\cdot \vec{B}_{ext})$ in $\delta H$. It is straightforward to show that for any constant vector $\vec{a}$, the matrix elements $\langle n,j,l,m_{j} \vert \gamma^{0}\vec{\gamma}\cdot \vec{a} \vert n,j,l',m_{j}'\rangle$ between the states of the same unperturbed energy vanish. Indeed, from the explicit form of the angular part of the matrix elements of the operator $\gamma^{0} \vec{\gamma}\cdot \vec{a}$ between the degenerate unperturbed wave functions, one can easily see that $\langle n,j,l,m_{j} \vert \gamma^{0}\vec{\gamma}\cdot \vec{a} \vert n,j,l',m_{j}'\rangle$ vanish for either $\vert m_{j}-m_{j}' \vert \neq 0$ or $1$, or $\vert l-l'\vert \neq 1$. However, when $\vert m_{j}-m_{j}' \vert =0 $ or $1$, and $\vert l -l' \vert =1$, the radial integral of $\langle n,j,l,m_{j} \vert \gamma^{0}\vec{\gamma}\cdot \vec{a} \vert n,j,l',m_{j}'\rangle$ vanishes. The constant term $-e \vec{v}\cdot \vec{B}_{ext}$ in $\delta H$ just shifts each energy level by the same amount. Therefore, we conclude that, in the presence of the uniform magnetic field $\vec{B}_{ext}$, the hydrogen spectrum is not altered by the LV $\delta H$ in first-order perturbation theory. 

We can also consider the change to the hydrogen energy levels in the presence of a uniform electric field $\vec{E}_{ext}$. Again, we assume that the external electric field $\vec{E}_{ext}$ is weak so that the unperturbed Hamiltonian is the Dirac Hamiltonian of the hydrogen atom. Besides the Stark effect which mixes the $2s$ and $2p$ states, we also need to calculate the matrix elements of the LV perturbation between the degenerate unperturbed states:  $-e \langle n,j,l\,', m_{j}'  \vert \gamma^{0} \vec{\gamma}\cdot (\vec{v} \times \vec{E}_{ext}) \vert n,j, l, m_{j} \rangle$. Since $\vec{v}\times \vec{E}_{ext}$ is a constant vector, we know that $-e \langle n,j,l\,', m_{j}'  \vert \gamma^{0} \vec{\gamma}\cdot (\vec{v} \times \vec{E}_{ext}) \vert n,j, l, m_{j} \rangle=0$ by the same reasoning as before, and therefore the interaction $-e \gamma^{0} \vec{\gamma}\cdot (\vec{v} \times \vec{E}_{ext})$ from the LV perturbation $\delta H$ does not add any new effect on the hydrogen energy levels.

\section{Model II}
We now turn to another model constructed from QED modified by another dimension-five LV operator $j^{\mu}v^{\nu}F_{\mu\nu}$, so that the Lagrange density is given by 
\begin{equation}
\mathcal{L}_{2}=-{1\over 4} F_{\mu\nu} F^{\mu\nu} + \overline{\Psi} \left( i \Dslash -m -e\gamma^{\mu}v^{\nu}  F_{\mu\nu}   \right) \Psi .  \label{L3}
\end{equation} 
The field equations which follows from~(\ref{L3}) are
\begin{equation}
\partial_{\nu} F^{\nu\mu}= \left( 1+v^{\nu}\partial_{\nu}\right) j^{\mu}.
\end{equation} 
In terms of components, we obtain
\begin{eqnarray}
\vec{\nabla}\cdot \vec{E}&&= \left(1+v^{0} {\partial \over \partial t} + \vec{v}\cdot \vec{\nabla} \right)\rho, \label{a1}\\
\vec{\nabla}\times \vec{B} - {\partial \vec{E} \over \partial t} &&=\left( 1+v^{0} {\partial \over \partial t} + \vec{v}\cdot \vec{\nabla} \right)\vec{j}\label{a2}.
\end{eqnarray} 
Using the continuity equation, we have
\begin{equation}
\left( v^{0}\partial_{0}+\vec{v}\cdot \vec{\nabla} \right)\rho=\vec{\nabla}\cdot\left(\rho \vec{v}-v^{0} \vec{j} \right)
\end{equation} 
and
\begin{equation}
\left( v^{0}\partial_{0}+\vec{v}\cdot \vec{\nabla} \right)\vec{j}=\vec{\nabla}\times\left( \vec{j}\times \vec{v}\right)-\partial_{0}     \left(\rho \vec{v}-v^{0} \vec{j} \right),
\end{equation} 
and thus the inhomogeneous Maxwell equations~(\ref{a1}) and~(\ref{a2}) can be expressed as
\begin{eqnarray}
\vec{\nabla}\cdot \vec{D}&&= \rho, \label{a3}\\
\vec{\nabla}\times \vec{H} - {\partial \vec{D} \over \partial t} &&=\vec{j}\label{a4},
\end{eqnarray} 
where the effective displacement field $\vec{D}$ and the effective magnetic field $\vec{H}$ are given, respectively, by
\begin{equation}
\vec{D}=\vec{E}+\left(v^{0}\vec{j}-\rho\vec{v}\right)\equiv \vec{E}+\overrightarrow{\tilde{P}},
\end{equation} 
and
\begin{equation}
\vec{H}=\vec{B}-\left(\vec{j}\times\vec{v}\right)\equiv \vec{B}-\overrightarrow{\tilde{M}}.
\end{equation} 
Compared with the model~(\ref{L1}) in Sec. II, we see that the effective polarization$\overrightarrow{\tilde{P}}$  and the effective magnetization $\overrightarrow{\tilde{M}}$ satisfy $\overrightarrow{\tilde{P}}=- \overrightarrow{M}$ and $\overrightarrow{\tilde{M}}=\overrightarrow{P}$.
This is not surprising, since the dimension-five operator $j^{\mu}v^{\nu}\tilde{F}_{\mu\nu}$ in model~(\ref{L1}) can be written as 
\begin{equation}
j^{\mu}v^{\nu}\tilde{F}_{\mu\nu} = {1\over 2} \epsilon_{\mu\nu\alpha\beta} M^{\mu\nu} F^{\alpha\beta}\equiv \tilde{M}_{\mu\nu}F^{\mu\nu},
\end{equation} 
and the duality between $(\overrightarrow{P},\overrightarrow{M})$ and $(\overrightarrow{\tilde{M}},-\overrightarrow{\tilde{P}} )$ follows immediately. Together with the homogeneous Maxwell equations, one can show that, in the presence of stationary sources, the gauge field $A^{\mu}$ is given by
\begin{equation}
A^{\mu}(\vec{r})={1\over 4\pi} \int d^{3}r' { (1+\vec{v}\cdot \vec{\nabla}')j^{\mu}(\vec{r}\,') \over \vert \vec{r} - \vec{r}\,' \vert}.\label{A}
\end{equation}
It follows from~(\ref{A}) that, different from the consequence of the model $\mathcal{L}_{1}$, a nonvanishing electric field cannot arise from neutral sources and a nonvanishing magnetic field cannot arise from steady and irrotational current density. 
 
The modified Dirac equation following from $\mathcal{L}_{2}$ reads
 \begin{equation}
\left(i \Dslash -m -e \gamma^{\mu} v^{\nu} F_{\mu\nu} \right)\Psi=0 .
\end{equation}
 Again, we can easily identify from the above equation the Hamiltonian operator $\tilde{H}$ of one-particle quantum mechanics:
 \begin{equation}
\tilde{H}=\gamma^{0} \left(\vec{\gamma}\cdot \vec{p} + e\Aslash + m - ev^{0} \vec{\gamma}\cdot \vec{E} +e \gamma^{0} \vec{v}\cdot \vec{E} -e \vec{\gamma}\cdot (\vec{v}\times\vec{B})  \right)\\
=H_{0} +\delta \tilde{H } \label{H2}
\end{equation}  
where the Dirac Hamiltonian $H_{0}$ is the same as before and $\delta \tilde{H}=-e(v^{0} \gamma^{0}\vec{\gamma}\cdot \vec{E} - \vec{v}\cdot \vec{E}+ \gamma^{0}\vec{\gamma}\cdot (\vec{v}\times\vec{B}))$ is the LV perturbation. We note that $\delta \tilde{H}$ can be obtained from $\delta H$ in~(\ref{H1}) by changing $\vec{E}\rightarrow \vec{B}$ and $\vec{B} \rightarrow -\vec{E}$. This is again due to the dual relation between the operator $j^{\mu}v^{\nu}\tilde{F}_{\mu\nu}$ in $\mathcal{L}_{1}$ and the operator $j^{\mu}v^{\nu}F_{\mu\nu}$ in $\mathcal{L}_{2}$.

To consider the first-order energy shift in the states of hydrogen atom induced by $\delta \tilde{H}$ in the absence of external fields, we need to calculate the following matrix elements:
 \begin{equation}
\langle n,j,l\,', m_{j}'  \vert \delta \tilde{H} \vert n,j, l, m_{j} \rangle=-e \langle n,j,l\,', m_{j}'  \vert (v^{0}\gamma^{0}\vec{\gamma}-\vec{v})\cdot\vec{E} \vert n,j, l, m_{j} \rangle \label{matrix}
\end{equation} 
where $\vec{E} $ is the Coulomb field. Once again, without loss of generality, we can choose a coordinate system in which the $z$ axis is in the direction of $\vec{v}$. Since it is easy to show that $J_{z}$ commutes with $(v^{0}\gamma^{0}\vec{\gamma}-\vec{v})\cdot\vec{E}$, by the same argument as given in Sec. II we know that the matrix elements~(\ref{matrix}) vanish for different states of the same unperturbed energy. As for the diagonal matrix elements, a straightforward calculation gives
\begin{eqnarray}
&-&e\langle   n,j,l,m_{j}\vert(v^{0}\gamma^{0}\vec{\gamma}-\vec{v})\cdot\vec{E} \vert n,j, l, m_{j} \rangle \nonumber\\
&=&\alpha \int dr d\cos{\theta}d{\varphi}\Big(-iF_{-}(\pm \kappa\vert r)\mathcal{Y}_{j,m_{j}}^{\dagger}(l,\frac{1}{2}\vert \hat{r}),\,F_{+}(\pm \kappa\vert r)\mathcal{Y}_{j,m_{j}}^{\dagger}(2j-l,\frac{1}{2}\vert \hat{r})\Big)\nonumber\\
&\,\,&[v^{0}\left(\begin{array}{cc}0
& \overrightarrow{\sigma} \\ \overrightarrow{\sigma} & 0
\end{array}\right)\cdot(\sin{\theta}\cos{\varphi}\hat{x}+\sin{\theta}\sin{\varphi}\hat{y}+\cos{\theta}\hat{z})-\vert \vec{v} \vert\cos{\theta}]
\binom{iF_{-}(\pm\kappa\vert r)\mathcal{Y}_{j,m_{j}}(l,\frac{1}{2}\vert \hat{r})}{F_{+}(\pm \kappa\vert r)\mathcal{Y}_{j,m_{j}}(2j-l,\frac{1}{2}\vert \hat{r})} \nonumber\\
&=&\alpha \int dr d\cos{\theta}d{\varphi}\{-iv^{0}F_{-}(\pm\kappa\vert r)F_{+}(\pm\kappa\vert r)(\sin{\theta}\cos{\varphi}\hat{x}+\sin{\theta}\sin{\varphi}\hat{y}+\cos{\theta}\hat{z})\cdot\nonumber\\
&\,\,&[\mathcal{Y}^{\dagger}_{j,m_{j}}(l,\frac{1}{2}\vert \hat{r})\overrightarrow{\sigma}\mathcal{Y}_{j,m_{j}}(2j-l,\frac{1}{2}\vert \hat{r})-\mathcal{Y}^{\dagger}_{j,m_{j}}(2j-l,\frac{1}{2}\vert \hat{r})\overrightarrow{\sigma}\mathcal{Y}_{j,m_{j}}(l,\frac{1}{2}\vert \hat{r})]-\vert \vec{v} \vert \cos{\theta}[\nonumber\\
&\,\,&F_{-}^{2}(\pm\kappa\vert r)\mathcal{Y}^{\dagger}_{j,m_{j}}(l,\frac{1}{2}\vert \hat{r})\mathcal{Y}_{j,m_{j}}(l,\frac{1}{2}\vert \hat{r})+F_{+}^{2}(\pm\kappa\vert r)\mathcal{Y}^{\dagger}_{j,m_{j}}(2j-l,\frac{1}{2}\vert \hat{r})\mathcal{Y}_{j,m_{j}}(2j-l,\frac{1}{2}\vert \hat{r})]\}\nonumber\\
&=&\mp i \alpha v^{0} \int dr d\cos{\theta}d{\varphi}\,F_{-}(\pm\kappa\vert r)F_{+}(\pm\kappa\vert r)(-1)^{j-l+\frac{1}{2}}\{\nonumber\\
&\,\,&[-\sqrt{\frac{(j-m_{j}+1)(j+m_{j})}{(2j)(2j+2)}}\cos{\theta}{Y^{*}}_{j+\frac{1}{2}}^{m_{j}-\frac{1}{2}}Y_{j-\frac{1}{2}}^{m_{j}-\frac{1}{2}}-\sqrt{\frac{(j-m_{j}+1)(j-m_{j})}{(2j)(2j+2)}}\sin{\theta}e^{-i\varphi}{Y^{*}}_{j+\frac{1}{2}}^{m_{j}-\frac{1}{2}}Y_{j-\frac{1}{2}}^{m_{j}+\frac{1}{2}}\nonumber\\
&\,\,&+\sqrt{\frac{(j+m_{j}+1)(j+m_{j})}{(2j)(2j+2)}}\sin{\theta}e^{i\varphi}{Y^{*}}_{j+\frac{1}{2}}^{m_{j}+\frac{1}{2}}Y_{j-\frac{1}{2}}^{m_{j}-\frac{1}{2}}-\sqrt{\frac{(j+m_{j}+1)(j-m_{j})}{(2j)(2j+2)}}\cos{\theta}{Y^{*}}_{j+\frac{1}{2}}^{m_{j}+\frac{1}{2}}Y_{j-\frac{1}{2}}^{m_{j}+\frac{1}{2}}]\nonumber\\
&\,\,&-[-\sqrt{\frac{(j-m_{j}+1)(j+m_{j})}{(2j)(2j+2)}}\cos{\theta}{Y^{*}}_{j-\frac{1}{2}}^{m_{j}-\frac{1}{2}}Y_{j+\frac{1}{2}}^{m_{j}-\frac{1}{2}}+\sqrt{\frac{(j+m_{j}+1)(j+m_{j})}{(2j)(2j+2)}}\sin{\theta}e^{-i\varphi}{Y^{*}}_{j-\frac{1}{2}}^{m_{j}-\frac{1}{2}}Y_{j+\frac{1}{2}}^{m_{j}+\frac{1}{2}}\nonumber\\
&\,\,&-\sqrt{\frac{(j-m_{j}+1)(j-m_{j})}{(2j)(2j+2)}}\sin{\theta}e^{i\varphi}{Y^{*}}_{j-\frac{1}{2}}^{m_{j}+\frac{1}{2}}Y_{j+\frac{1}{2}}^{m_{j}-\frac{1}{2}}-\sqrt{\frac{(j+m_{j}+1)(j-m_{j})}{(2j)(2j+2)}}\cos{\theta}{Y^{*}}_{j-\frac{1}{2}}^{m_{j}+\frac{1}{2}}Y_{j+\frac{1}{2}}^{m_{j}+\frac{1}{2}}]\}\nonumber\\
&=&0
\end{eqnarray}
for $l=j\pm{1\over 2}$, in which we have used the facts that terms proportional to $\vert \vec{v} \vert$ are odd functions of $\cos{\theta}$ and terms proportional to $v^{0}$ cancel each other out. Thus the energy levels of the hydrogen atom are not shifted by the LV perturbation $\delta \tilde{H}$.

The shift of the hydrogen energy levels induced by $\delta \tilde{H}$ in the presence of uniform external fields can be easily analyzed in the same way as we did in Sec. II, assuming the external fields (denoted again by $\vec{E}_{ext}$ and $\vec{B}_{ext}$) are weak. Since $v^{0}\vec{E}_{ext}$ and $\vec{v}\times \vec{B}_{ext}$ are constant vectors, using the fact that for any constant vector $\vec{a}$ the matrix elements $\langle n,j,l\,', m_{j}'  \vert \gamma^{0}\vec{\gamma}\cdot \vec{a} \vert n,j, l, m_{j} \rangle$ vanish, and knowing that the interaction $e\vec{v}\cdot \vec{E}_{ext}$ shifts each energy level by the same amount, we can conclude that, in the presence of uniform external fields, the LV perturbation $\delta \tilde{H}$ still produces no effect on the hydrogen spectrum in first-order perturbation theory.

\section{Conclusion}
In this paper, QED modified by dimension-five LV operators $j^{\mu}v^{\nu}\tilde{F}_{\mu\nu}$ and $j^{\mu}v^{\nu}F_{\mu\nu}$ has been studied separately. In both cases, we have identified the effective polarization and magnetization, which are components of the rank-2 object $j^{[\mu}v^{\nu]}$, from the field equations of motion. In particular, we find that, with the LV interaction $j^{\mu}v^{\nu}\tilde{F}_{\mu\nu}$, any charged spinor has a spin-independent magnetic dipole moment density $\rho \vec{v}$, along with the one associated with its spin. Also, a static electric field can arise from stationary and neutral sources. These novel properties do not come up from the other interaction $j^{\mu}v^{\nu}F_{\mu\nu}$.

We have computed the shift in the energies of the states of a hydrogen atom in first-order perturbation theory. Our result shows that only the $CP$-even operator $j^{i}v^{k} \tilde{F}_{ik}$ produces the energy shifts, given by~(\ref{dE}), and the degeneracy of each level is completely removed. Interestingly, the breakdown of Lorentz boost symmetry, induced by the $v^{0}$ terms, in these two models plays no role in determining the atomic energy spectrum. In the presence of uniform external fields, both LV interactions add no new effect on the hydrogen spectrum. 
 
It would be interesting to study the loop corrections to these two models and make physical predictions in the framework of effective field theories. The analysis will be reported elsewhere.

\begin{acknowledgments}
This research was supported in part by the National Nature Science Foundation of China under Grant No. 10805024.
\end{acknowledgments}


\end{document}